\documentclass[runningheads]{llncs}
\usepackage{amsmath,amssymb,amsfonts}
\usepackage{graphicx}
\usepackage{lmodern,array,booktabs}
\usepackage{bbding}
\usepackage{comment}
\usepackage[colorlinks=true, allcolors=blue]{hyperref}
\usepackage{todonotes}
\begin{document}
\title{Attention-Enhanced Hybrid Feature Aggregation Network for 3D Brain Tumor Segmentation}
\titlerunning{Attention-Enhanced Hybrid Feature Aggregation Network}
% If the paper title is too long for the running head, you can set
% an abbreviated paper title here
%
    \author{Ziya Ata Yazıcı\inst{1}\Envelope \orcidID{0000-0001-7051-833X} \and
İlkay Öksüz\inst{1}\orcidID{0000-0001-6478-0534} \and
Hazım Kemal Ekenel\inst{1,2}\orcidID{0000-0003-3697-8548}}
\authorrunning{Yazıcı et al.}
% First names are abbreviated in the running head.
% If there are more than two authors, 'et al.' is used.
%
\institute{Istanbul Technical University, Department of Computer Engineering\\Istanbul, Turkey\\
\email{\{yaziciz21, oksuzilkay, ekenel\}@itu.edu.tr}\\ \and
Qatar University, Department of Computer Science and Engineering\\ Doha, Qatar\\
\email{hekenel@qu.edu.qa}}
\maketitle              % typeset the header of the contribution
\begin{abstract}
Glioblastoma is a highly aggressive and malignant brain tumor type that requires early diagnosis and prompt intervention. Due to its heterogeneity in appearance, developing automated detection approaches is challenging. To address this challenge, Artificial Intelligence (AI)-driven approaches in healthcare have generated interest in efficiently diagnosing and evaluating brain tumors. The Brain Tumor Segmentation Challenge (BraTS) is a platform for developing and assessing automated techniques for tumor analysis using high-quality, clinically acquired MRI data. In our approach, we utilized a multi-scale, attention-guided and hybrid U-Net-shaped model -- GLIMS -- to perform 3D brain tumor segmentation in three regions: Enhancing Tumor (ET), Tumor Core (TC), and Whole Tumor (WT). The multi-scale feature extraction provides better contextual feature aggregation in high resolutions and the Swin Transformer blocks improve the global feature extraction at deeper levels of the model. The segmentation mask generation in the decoder branch is guided by the attention-refined features gathered from the encoder branch to enhance the important attributes. Moreover, hierarchical supervision is used to train the model efficiently. Our model's performance on the validation set resulted in 92.19, 87.75, and 83.18 Dice Scores and 89.09, 84.67, and 82.15 Lesion-wise Dice Scores in WT, TC, and ET, respectively. The code is publicly available at \url{https://github.com/yaziciz/GLIMS}.

\keywords{Brain Tumor Segmentation \and Vision Transformer \and Deep Learning \and Hybrid \and Attention \and BraTS}
\end{abstract}

\newpage
\section{Introduction}
Glioblastoma is a type of brain tumor that falls under high-grade gliomas (HGG), which are aggressive and malignant tumors originating from brain glial cells. These tumors proliferate rapidly and often require surgery, radiotherapy, and have a poor prognosis in terms of survival \cite{glioblastoma}. Magnetic Resonance Imaging (MRI) has emerged as a crucial diagnostic tool for brain tumor analysis, providing detailed information about tumor location, size, and morphology. To comprehensively evaluate glioblastoma, multiple complimentary 3D MRI modalities, including T1, T1 with contrast agent (T1c), T2, and Fluid-attenuated Inversion Recovery (FLAIR), are utilized to highlight different tissue properties and areas of tumor spread \cite{van2019perfusion}. With the advent of AI in healthcare, there is an increasing demand for AI-driven intervention strategies in diagnosing and preliminary evaluating brain tumors from MRI scans. The accurate segmentation and characterization of glioblastoma using AI techniques has the potential to significantly improve treatment planning and patient outcome predictions. In medical imaging research, the BraTS challenge promotes innovation and collaboration in tumor segmentation. The challenge provides high-quality, clinically-acquired, 3D multimodal and multi-site MRI scans with their ground truth masks annotated by radiologists \cite{baid2021rsna}.

The hybrid approaches in medical image segmentation tasks have been previously proposed \cite{chen2022mixformer,yuan2023effective,bao2023hybrid,oktay2022attention}. These approaches involve integrating transformers, attention modules and convolutional layers to leverage the advantages of these structures; however, their implementation on the brain tumor segmentation task is limited. The utilization of Vision Transformer \cite{dosovitskiy2020image} (ViT) models, a sequence-to-sequence feature extractor, has greatly improved medical image segmentation tasks \cite{hatamizadeh2022swin,heidari2023hiformer}. These models have demonstrated their advantages over Convolutional Neural Network (CNN)-based models in terms of their global feature extraction ability and segmentation performance when a large number of available data exists. On the contrary, CNN models excel in extracting local features, which is particularly advantageous in region-based segmentation tasks, where overlapping regions require clear edge segmentation.

\begin{figure}[!h]
    \centering
    \includegraphics[width=340px]{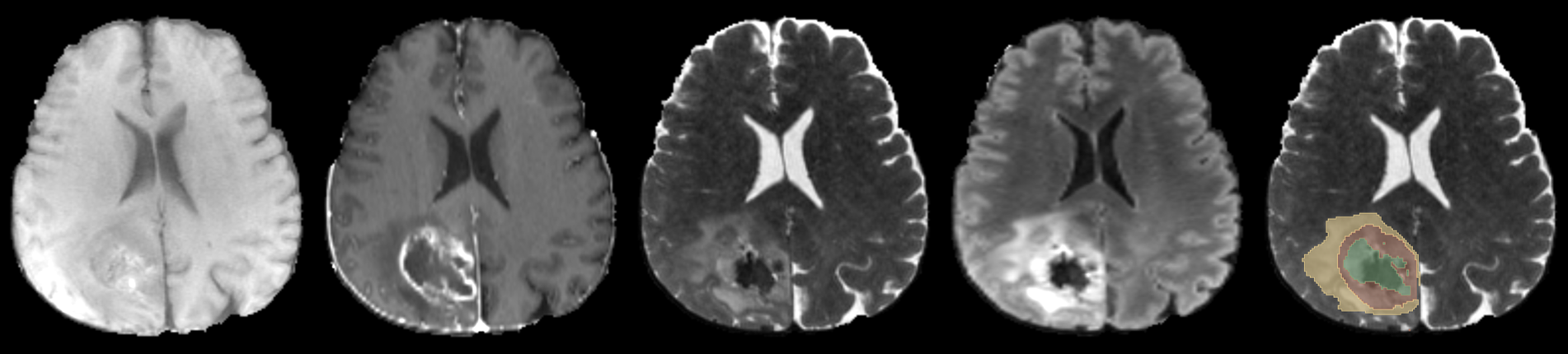}
    \caption{A sample MRI scan displayed in four modalities -- T1, T1c, T2, FLAIR -- and the corresponding segmentation mask, left to right. NCR is represented by green, ET by red, and ED by yellow.}
    \label{fig:Dataset}
\end{figure}

With this motivation, we propose a U-Net-shaped \cite{ronneberger2015u}  Attention-\textbf{G}uided \textbf{LI}ghtweight \textbf{M}ulti-\textbf{S}cale Hybrid Network (GLIMS) for 3D brain tumor segmentation, encompassing depth-wise multi-scale feature aggregation modules in a transformer-enhanced network. To improve the fine-grained segmentation mask prediction, we refine the encoder features via the channel and spatial-wise attention blocks as guidance on a skip connection. Furthermore, the model is supervised with multi-scale segmentation outputs, including the deeper decoder levels. With this approach, we participated in the Adult Glioblastoma Segmentation Task (Task 1) of the BraTS 2023 challenge, and our implementation ranked within the top 5 best-performing approaches in the validation phase.

\section{Dataset}

The dataset provided in BraTS 2023 consists of 1,251 multi-institutional 3D brain MRI scans in four modalities -- T1, T1c, T2, and FLAIR -- with the tumor segmented masks in four regions -- necrotic tumor core (NCR), peritumoral edematous tissue (ED), enhancing tumor (ET) and the background (Figure \ref{fig:Dataset}) \cite{karargyris2023federated,baid2021rsna,menze2014multimodal,bakas2017segmentation,bakas2017segmentation2}. The cross-sectional images of each modality are properly registered and the skull is removed from the images. The slices have a high-resolution isotropic voxel size of 1 x 1 x 1 mm$^3$, and each MRI scan has a size of 240 x 240 x 155 voxels in height, width, and depth. To comply with the ranking rules of the challenge, the given mask labels were converted into new label groups: Whole Tumor (WT) (NCR + ED + ET), Tumor Core (TC) (NCR + ET), and Enhancing Tumor (ET). A validation set with 219 cases without ground truth labels is also provided to evaluate the model performances through the official servers of BraTS 2023. 

\section{Methods}

In the following sections, the architecture of GLIMS, pre- and post-processing approaches, the deep supervision technique, the evaluation metrics and the implementation details are given.

\subsection{Model Overview}

\begin{figure}
    \centering
    \includegraphics[width=\linewidth]{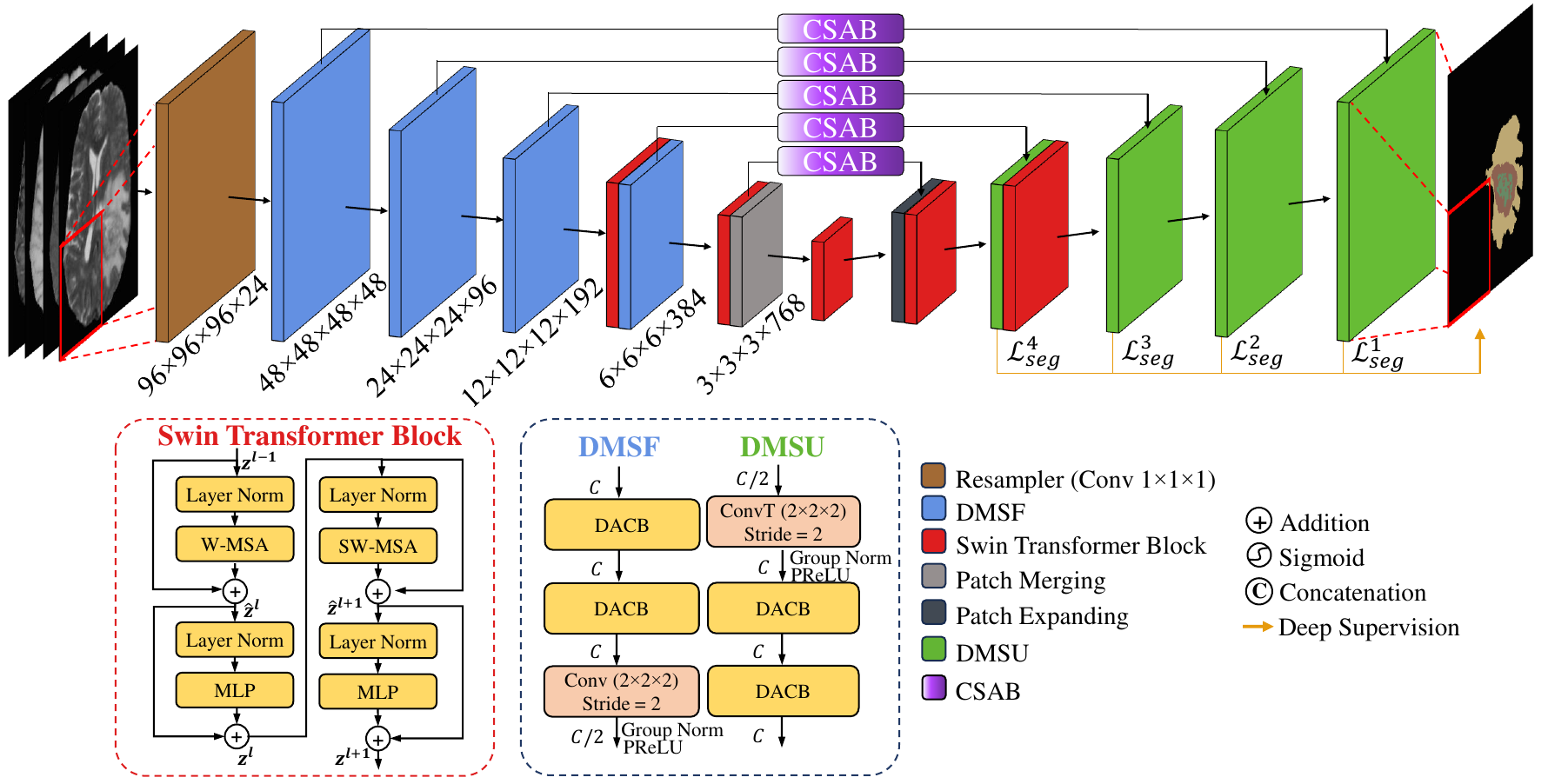}
    \caption{The proposed architecture of 3D segmentation model, GLIMS. Each color represents a unique module.}
    \label{fig:Model1}
\end{figure}

\begin{figure}[!t]
    \centering
    \includegraphics[width=\linewidth]{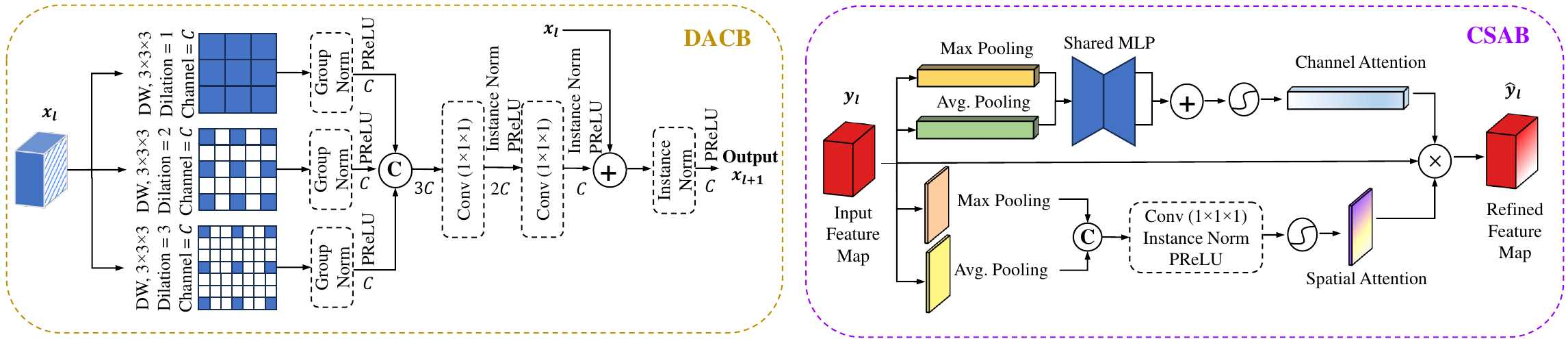}
    \caption{The proposed DACB and CSAB modules from left to right, respectively.}
    \label{fig:Model2}
\end{figure}

Our model's overall architecture is illustrated in Figure \ref{fig:Model1}, which utilizes \textbf{D}epth-Wise \textbf{M}ulti-\textbf{S}cale \textbf{F}eature Extraction (DMSF) modules and \textbf{D}epth-Wise \textbf{M}ulti-\textbf{S}cale \textbf{U}psampling (DMSU) modules in encoder and decoder branches, respectively. In each module, two consecutive \textbf{D}ilated Feature \textbf{A}ggregator \textbf{C}onvolutional \textbf{B}locks (DACB) are located. Depending on the branch, the convolutional blocks are followed with dilated $2 \times 2 \times 2$ convolution layer to downsample or transposed convolution to upsample. Each dilated convolutional layer in DACB is concatenated together, and two  $1 \times 1 \times 1$ point-wise convolutions are applied sequentially to weight further the important features more and reduce the channels gradually, as shown in Figure \ref{fig:Model2}. The resulting output is added to the input scan for the next layer to prevent gradient-vanishing. By this proposed module, the fine-grained features of the regions could be extracted in different resolutions, which provides robustness in both local and global feature extraction compared to the standalone convolution and transformer networks. The lower levels of the proposed model are designed as a hybrid combination of convolutions and transformer blocks to enhance the contextual and global feature extraction together. The main motivation behind the hybrid design was to utilize the locality of convolutions and globality of the transformer layers to benefit both overall and region-wise tumor segmentation. The Swin Transformer layers were used in the deeper layers, which utilize the shifted-window self-attention approach to reduce the trainable parameters and, therefore, the model's complexity. Finally, the refined features via the \textbf{C}hannel and \textbf{S}patial-Wise \textbf{A}ttention \textbf{B}locks (CSAB) (Figure \ref{fig:Model2}) from the encoder branch were fused to the decoder branch with skip connections. The CSAB module refines input feature maps, $y_l$, by selectively enhancing or inhibiting features in the channel and spatial dimensions separately. After obtaining the refined features, $\hat{y}_l$, the decoder leverages them to guide the mask predictions. 

The proposed model was designed to work efficiently on small graphical processing units due to its patch-based nature. A random patch from the whole input scan is sampled and processed with the model in each iteration. By this method, the training process requires less memory and benefits from a random-sampling augmentation process. Accordingly, the input size of the model is selected as $X \in \mathbb{R}^{H \times W \times D \times S}$, where $H$, $W$, and $D$ are chosen as 96. The initial input is resampled with a $1 \times 1 \times 1$ point-wise convolutional layer to have a depth of $S$ as 24. In each layer of the encoder branch, the spatial resolution of the feature matrix is halved and the channel resolution is doubled. The Swin Transformer block is used in the network's deeper encoder, decoder, and bottleneck parts for a hybrid approach. The input features to the transformer blocks are first partitioned with a patch size of $2 \times 2 \times 2$ to create tokens of $\left [ \frac{H}{2} \right ] \times \left [ \frac{W}{2} \right ] \times \left [ \frac{D}{2} \right ]$. The created patches are added with learnable positional embeddings in the shape of $\left [ \frac{H}{2} \right ] \times \left [ \frac{W}{2} \right ] \times \left [ \frac{D}{2} \right ] \times C$, where $C$ is the hidden size of the current layer. Self-attention modules are applied to non-overlapping embedding windows for efficient processing. To perform the attention at transformer level $l$, we equally partition 3D tokens into $\left [ \frac{H'}{M} \right ] \times \left [ \frac{W'}{M} \right ] \times \left [ \frac{D'}{M} \right ]$, where $M \times M \times M$ is the window resolution; $H'$, $W'$ and $D'$ are the current shape of the feature matrix in height, width, and depth, respectively. In the following layer $l+1$, the patches are shifted to capture local context and improve the model's ability to capture fine-grained local details. By shifting the patches, each patch can attend to its neighboring patches, allowing it to gather information from the surrounding local context. The shifting operation ensures that the receptive fields of the patches overlap, enabling the model to integrate the local feature relations effectively. To achieve this, the windows are shifted by $\left (\left [ \frac{M}{2} \right ], \left [ \frac{M}{2} \right ], \left [ \frac{M}{2} \right ]\right )$ voxels. The outputs of layers $l$ and $l+1$ are found as shown in Equation~\ref{eq:Swin}.

\begin{equation}
\begin{aligned}
\hat{z}_l &= \text{W-MSA}(\text{LN}(z_{l-1})) + z_{l-1} \\
z_l &= \text{MLP}(\text{LN}(\hat{z}_l)) + \hat{z}_l \\
\hat{z}_{l+1} &= \text{SW-MSA}(\text{LN}(z_l)) + z_l \\
z_{l+1} &= \text{MLP}(\text{LN}(\hat{z}_{l+1})) + \hat{z}_{l+1}
\end{aligned}
\label{eq:Swin}
\end{equation}

In Equation \ref{eq:Swin}, W-MSA, SW-MSA, LN, and MLP represent windowed and shifted window multi-head self-attention modules, layer normalization, and multi-layer perceptron, respectively. The patches are shifted after every W-MSA layer using cyclic-shift \cite{liu2021swin}. This ensures that the number of windows for self-attention remains the same and the complexity does not increase. Finally, GLIMS has 72.30G FLOPs and 47.16M trainable parameters, making it comparably lightweight to the previous studies.

%ADD MODEL CONFIG AS SWIN UNETR, embed dim etc., feature size etc...
\newpage
\subsection{Data Pre-processing}

Each MRI scan has a NIfTI format with separate modalities and a segmentation mask in three classes. In the scope of the challenge, the segmentation results should be evaluated in modified sub-regions such as WT, TC, and ET. Therefore, the label modifications and augmentations were applied on the fly during training by using the MONAI \cite{monai2020monai} framework. To implement the patch-based training technique, a randomly cropped volume with a size of $96 \times 96 \times 96$ is taken from a 3D MRI scan. Each cropped region was flipped in $X$-$Y$-$Z$ axes with an equal probability of 0.5. Z-normalization was applied to the scans and each normalization was performed independently between the modalities. The normalized intensities were scaled and shifted to simulate the different scanner properties with a factor of 0.2 and a probability of 0.2. To make the model robustly generalize the unseen data, the contrast of the cropped volumes was changed with a gamma value between $[0.5, 4.5]$ and a probability of 0.2. Additionally, Gaussian noise was added with $\sigma = 0.2$, $\mu = 0$, and Gaussian smoothing was applied with a varying $\sigma$ value in $X$-$Y$-$Z$ axes between $[0.25, 1.15]$ with a probability of 0.2. %During the validation process, the scan patches were sampled with an overlap of 0.8.

\subsection{Evaluation Metrics \& Loss Function}

The segmentation performance of the models was evaluated with Dice Score and Hausdorff95 (HD95) distance metrics. Compared to the previous BraTS challenges, two new evaluation metrics were introduced this year: Lesion-wise Dice Score and Lesion-wise HD95. These metrics provide insights into how well models detect and segment multiple individual lesions within a scan, addressing the importance of identifying large and small lesions in clinical practice. For the Lesion-wise metrics, the ground truth masks undergo a $3 \times 3 \times 3$ mm$^3$ dilation before calculating the Dice Score and HD95. Following the process, connected component analysis is performed on the predictions to compare the lesions with the ground truth labels by counting the number of True Positive (TP), False Positive (FP), True Negative (TN), and False Negative (FN) predicted voxels.

The models were optimized with the combination of Dice Loss (Equation \ref{eq:DiceLoss}) and Cross-Entropy Loss (Equation \ref{eq:CrossEntropy}) as shown in Equation \ref{eq:LossTotal}.

\begin{equation}
    \mathcal{L}_{Dice} = \frac{2}{K}\sum_{k=1}^{K}\frac{\sum_{i=1}^{N} y_{i,k} p_{i,k}}{\sum_{i=1}^{N} y_{i,k}^2 + \sum_{i=1}^{N} p_{i,k}^2}
\label{eq:DiceLoss}
\end{equation}

\begin{equation}
    \mathcal{L}_{CE} =\sum_{k=1}^{K}\sum_{i=1}^{N} y_{i,k} \log(p_{i,k})
\label{eq:CrossEntropy}
\end{equation}

\begin{equation}
    \mathcal{L}_{Seg} =1 - \alpha \mathcal{L}_{Dice} - \beta \mathcal{L}_{CE}
\label{eq:LossTotal}
\end{equation}
where $K$ represents the total number of classes, $N$ represents the number of voxels, $y$ refers to the ground truth labels, and $p$ refers to the predicted one-hot classes. The weights of $\alpha$ and $\beta$ were selected as 0.5 to calculate the total loss.

\subsection{Deep Supervision}

Deep supervision \cite{zhu2017deeply} is a technique of computing the loss function, $\mathcal{L}_{DS}$, from the last layer and incorporating the deeper layers of the decoder. It involves training CNNs with multiple intermediate supervision signals, allowing for better performance and improved segmentation results. Traditionally, the network is trained end-to-end with a single mask output, making it difficult to identify and correct errors at different stages of the network. However, by introducing intermediate supervision, additional loss functions are applied at multiple network layers, enabling the network to learn more discriminative and informative features. The utilized loss function can be seen in Equation \ref{eq:DS}, where each $L^{i}_{seg}, i \in \{1,2,3,4\}$ represents the loss values corresponding to the combination of $\mathcal{L}_{Dice}$ and $\mathcal{L}_{CE}$ for level $i$. While shallower layers have the highest weight, the given weight decreases for the deeper layers.

\begin{equation}
    \mathcal{L}_{DS} = \mathcal{L}^1_{seg} + \frac{1}{2}\mathcal{L}^2_{seg} \\
    + \frac{1}{4}\mathcal{L}^3_{seg} + \frac{1}{8}\mathcal{L}^4_{seg}
\label{eq:DS}
\end{equation}

\subsection{Post-processing}

 The post-processing of the predicted region masks could improve the metric results significantly, as experimented by the previous studies \cite{kotowski2021coupling,jabareen2021segmenting}. Especially in the cases where no ground truth class occurs in a specific slice, eliminating false positive predictions increases the Dice Score from 0 to 100. Thus, to advance the model's performance more, three post-processing steps were considered to be applied in our approach:

\begin{itemize}
    \item \textbf{Region Removal:} Small regions with $\#$ of voxels less than $\psi$ with mean confidence less than $\theta$ are removed from the predictions. This is applied to eliminate the false positive predictions of the scans with no ground truth label to increase the Dice Score from 0 to 100.
    \item \textbf{Threshold Modification:} The models are optimized to perform thresholding at 0.5 while selecting the hard labels from the region probabilities. However, adjusting the confidence level during inference could be beneficial to eliminate exceeding region borders or including border voxels to the region.
    \item \textbf{Center Filling:} To improve the segmentation performance of class ET and TC, the center voxels of the ET components are replaced with NCR. This could improve ET and TC Dice Scores.
\end{itemize}

\subsection{Implementation Details}

Our model, GLIMS, was implemented in PyTorch framework v2.0.1 by using the MONAI library v1.2.0. The experiments were performed on a single NVIDIA 3090 GPU with 24 GB VRAM for 800 epochs in a 5-fold cross-validation approach. The learning rate was set to 0.001 and the cosine annealing scheduler was used to update the learning rate. The parameters were updated with the AdamW optimizer. The batch size was selected as two, and a sliding window approach with a 0.8 inference overlap was applied using $96 \times 96 \times 96$ patch size. The model parameters were saved for the highest Dice Scores on the internal validation set, and the experiments were performed on the best model states.

\section{Results}

The performance of the proposed model was compared with the nnU-net \cite{isensee2021nnu} architecture as a baseline, which was among the top-performing models of the previous years. %We selected the nnU-Net as the baseline model to evaluate our model's performance improvement. 
The experiments were conducted using the same training dataset and data distribution for both models. We maintained consistency in all implementation details while conducting the experiments. The results in Table \ref{tab:TrainingResults} show that our method performed better by 0.88\% in the overall performance in terms of the Lesion-wise Dice Score.  %; however, to finalize the architectural design of the proposed model, the ablation studies were performed on a single fold with deterministic sessions, trained for 100 epochs. However, for the 5-fold cross-validation experiments, the models were trained for 800 epochs.

\begin{table}[]
\centering
\renewcommand{\arraystretch}{1}
\caption{The experimental results of 5-fold cross-validation in Legacy Dice Scores (\%) without post-processing is applied.}
\label{tab:TrainingResults}
\resizebox{300pt}{!}{%
\addtolength\tabcolsep{2pt}
\begin{tabular}{|c|ccccc|c|}
\hline
Model   & Fold 1          & Fold 2          & Fold 3          & Fold 4          & Fold 5           & Average $\uparrow$         \\ \hline
nnU-net \cite{isensee2021nnu} & 90.12          & 90.80          & 91.45          & 91.75          & 90.42           & 90.91          \\
\textbf{Ours}    & \textbf{91.19} & \textbf{91.52} & \textbf{92.74} & \textbf{92.21} & \textbf{91.27} & \textbf{91.79} \\ \hline
\end{tabular}
}
\vspace{-5pt}
\end{table}

The experiments were extended to the validation set to select the best-performing settings. These studies cover post-processing and ensemble approaches with varying parameter selection. We first observed the influence of the post-processing techniques on the validation set performance. The results in Table \ref{tab:Validation} show the improvement of the methods as they were applied to the predicted segmentation masks. As the post-processing techniques were applied, the average Lesion-wise Dice Score improvement was observed as 15.5\% for Fold 0. The removal of small false positive predictions in the slices without true positive ground truth labels increased the individual Dice Scores from 0 to 100. 

\begin{figure} [!h]
    \centering
    \includegraphics[width=240px]{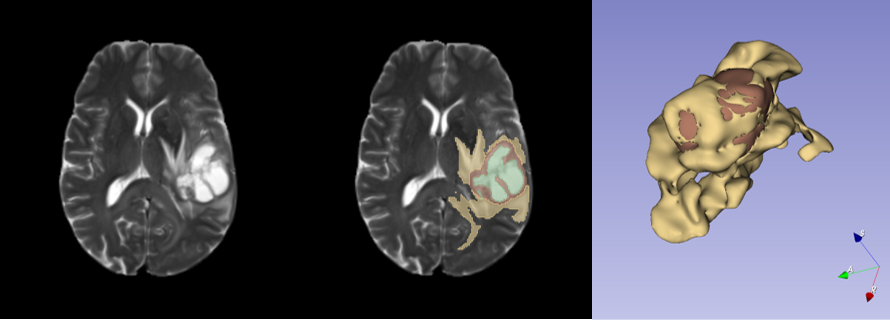}
    \caption{The prediction result of Case ID: 208 in the validation set. Left: The T2 image of the slice. Middle: The segmented output. Right: 3D rendered visualization of the tumor. The yellow, red, and green colors represent ED, ET, and NCR regions.}
    \label{fig:Result}
\end{figure}

\begin{table}
\centering
\centering
\caption{The experimental results on the online validation set with different post-processing and ensemble approaches. The results are given in the Lesion-wise metrics and were obtained through the submission system of BraTS 2023. \textit{RR}: Region Removal, \textit{TM}: Threshold Modification, \textit{CF}: Center Filling. $^\dagger$The TC threshold was set to 0.6. *The ET threshold was set to 0.6.}
\label{tab:Validation}
\resizebox{\linewidth}{!}{%
\begin{tabular}{|l|c|c|c|c|c|c|c|c|c|c|} 
\hline
\textbf{Model} & \multicolumn{3}{c|}{\textbf{Method}} & \multicolumn{3}{c|}{\textbf{HD95 (mm) $\downarrow$}} & \multicolumn{4}{c|}{\textbf{Dice Score (\%) $\uparrow$}} \\ 
\cline{2-11}
 & RR & TM & CF & ET & TC & WT & ET & TC & WT & Avg. \\ 
\cline{1-11}
Fold 0 &  &  &  & 91.49 & 70.59 & 107.07 & 66.66 & 73.13 & 66.20 & 68.66 \\
Fold 1 &  &  &  & 141.51 & 112.12 & 133.34 & 54.56 & 62.32 & 59.38 & 58.75 \\
Fold 0 & \checkmark &  &  & 25.63 & 33.10 & 19.07 & 81.85 & 81.76 & 87.98 & 83.86 \\
Fold 0$^\dagger$ & \checkmark & \checkmark &  & 25.63 & 29.42 & 19.07 & 81.85 & 82.57 & 87.98 & 84.13 \\
Fold 0$^\dagger$ & \checkmark & \checkmark & \checkmark & 25.63 & 29.42 & 18.43 & 81.85 & 82.57 & 88.12 & 84.18 \\
Fold 2$^\dagger$ & \checkmark & \checkmark & \checkmark & 26.34 & 23.75 & 17.35 & 82.04 & 83.91 & 88.56 & 84.84 \\
Fold 3$^\dagger$ & \checkmark & \checkmark & \checkmark & 28.76 & 34.30 & 16.47 & 80.91 & 80.87 & 88.49 & 83.42 \\
Fold 2+4$^{*\dagger}$ & \checkmark & \checkmark & \checkmark & \textbf{25.16} & \textbf{20.75} & \textbf{15.80} & \textbf{82.15} & \textbf{84.67} & \textbf{89.09} & \textbf{85.30} \\
\hline
\end{tabular}
}
\end{table}

Based on the experiments, the best settings were selected as the removal of the ET and NCR regions smaller than $75\text{mm}^3$ and ED regions smaller than $500\text{mm}^3$ if the model's average confidence is less than 0.9. For the threshold modification method, threshold levels between 0.5 and 0.7 were tested. It was observed that changing the threshold of ET and TC from 0.5 to 0.6 was the most effective approach, which reduced the false positive TC predictions. Lastly, after a connected component analysis in 3D, the voxels inside the ET regions were replaced as NCR voxels. This approach improved the TC segmentation performance as well as WT if any unassigned voxels occurred. The ensemble methods also improved the results; thus, our submission was based on the combination of post-processed Fold 2 and Fold 4 models, as it yielded the highest Lesion-wise Dice Score. Lastly, as a qualitative result of the approach, a sample mask prediction and a 3D-rendered tumor output from the validation set can be seen in Figure \ref{fig:Result}.

% Please add the following required packages to your document preamble:
% \usepackage{graphicx}
\begin{table}[]
\centering
\caption{The segmentation performance of the proposed approach in Lesion-wise metrics on the online validation and test sets. The scores are retrieved from the official submission system of BraTS 2023.}
\label{tab:TestingResults}
\resizebox{330pt}{!}{%
\addtolength\tabcolsep{2pt}
\begin{tabular}{|c|cccc|cccc|}
\hline
\textbf{Data Split} & \multicolumn{4}{c|}{\textbf{HD95 (mm) $\downarrow$}}                                         & \multicolumn{4}{c|}{\textbf{Dice Score (\%) $\uparrow$}}                                     \\ \cline{2-9} 
                    & \multicolumn{1}{c|}{ET}    & \multicolumn{1}{c|}{TC}    & \multicolumn{1}{c|}{WT}    & Avg.  & \multicolumn{1}{c|}{ET}    & \multicolumn{1}{c|}{TC}    & \multicolumn{1}{c|}{WT}    & Avg.  \\ \hline
Validation          & \multicolumn{1}{c|}{25.16} & \multicolumn{1}{c|}{20.75} & \multicolumn{1}{c|}{15.80} & 20.57 & \multicolumn{1}{c|}{82.15} & \multicolumn{1}{c|}{84.67} & \multicolumn{1}{c|}{89.09} & 85.30 \\
Test                & \multicolumn{1}{c|}{26.01} & \multicolumn{1}{c|}{34.68} & \multicolumn{1}{c|}{28.50} & 29.73 & \multicolumn{1}{c|}{83.67} & \multicolumn{1}{c|}{82.90} & \multicolumn{1}{c|}{85.97} & 84.18 \\ \hline
\end{tabular}%
}
\end{table}

The approach that yielded the best validation result was also evaluated on the blinded test set. Compared to the validation split, the MRI samples in the testing set are sampled from a different patient cohort and multi-institutional sensors compared to the training set. According to the post-challenge results on the testing data, our model had a slight decrease in the mean Lesion-wise Dice Score and an increase in Lesion-wise HD95 metrics by 1.12\% and 9.16 mm, achieving an average of 84.18\% Dice Score and 29.73 mm HD95 in lesion-wise performance,  as shown in Table \ref{tab:TestingResults}.

\section{Discussion and Conclusions}

In this study, we proposed a U-net-shaped hybrid 3D MRI segmentation model for the BraTS 2023 challenge. We utilized depth-wise multi-scale feature extraction blocks and attention modules to perform fine-grained region-based segmentation tasks with high Lesion-wise performance. To reduce the number of trainable parameters; transformer blocks were incorporated in the bottleneck, the convolutional layers were converted to perform depth-wise operations and the sliding window inference technique was used. An attention guidance method was implemented to support tumor region prediction by utilizing important features from the encoder branch. Additionally, the impact of the post-processing techniques on the segmentation performance was examined. Although the Legacy Dice Score was less affected by post-processing techniques; removing small false positive regions from the outputs, adjusting the prediction threshold, and filling the center of the connected components significantly improved the Lesion-wise Dice Score. On the online validation set, GLIMS achieved a Lesion-wise Dice Score of 0.8909, 0.8467, and 0.8215 for WT, TC, and ET classes, respectively, placing it among the top 5 best-performing approaches in the validation phase. In the testing phase, as the data distribution changed compared to the training set, our approach achieved 84.18\% Lesion-wise Dice Score by a decrease of 1.12\%, and 29.73 mm Lesion-wise HD95 by an increase of 9.16 mm compared to the validation result.

The results represent our model's enhanced performance on the 3D brain tumor segmentation task and the robustness of the post-processing techniques. As a slight performance decrease occurred in the test set, we could diversify the representation of the patients and the sensors in the training dataset by synthetically generating healthy and diseased MRI scans. Therefore, as a further study, synthetic data generation techniques could be employed to improve the model's generalizability on unseen data by introducing MRI samples in wider settings. Additionally, to reduce the possible defects in the predicted masks further, new post-processing methods could be employed by integrating the field knowledge of the physicians. Moreover, the proposed models should be further optimized to run efficiently on the end-user side. Although increasing the model size generally improves the segmentation performance, it becomes challenging to deploy and use effectively. Thus, in the future, we aim to investigate the impact of the synthetic data, reduce the model complexity more by utilizing lightweight yet robust modules, and perform better optimization techniques.

%
% ---- Bibliography ----
%
% BibTeX users should specify bibliography style 'splncs04'.
% References will then be sorted and formatted in the correct style.
%

\subsubsection*{Acknowledgements.} This study has been partially funded by Istanbul Technical University, Department of Computer Engineering and Turkcell via a Research Scholarship grant provided to Ziya Ata Yazıcı.

\bibliographystyle{splncs04}
\bibliography{bib}

\end{document}